# An Illustrated History of Black Hole Imaging : Personal Recollections (1972-2002)

Jean-Pierre Luminet

*Aix-Marseille Université, CNRS, Laboratoire d'Astrophysique de Marseille (LAM) UMR 7326*
*& Centre de Physique Théorique de Marseille (CPT) UMR 7332*
*& Observatoire de Paris (LUTH) UMR 8102*
*France*
*E-mail:* jean-pierre.luminet@lam.fr

## Abstract

The Event Horizon Telescope Consortium is on the verge to provide the first telescopic image of massive black holes SgrA* and M87* surrounded by accretion disks, at a resolution scale comparable to the size of their event horizons. Well before this remarkable achievement made possible by VLBI radio astronomy, many researchers used the computer to reconstruct how a black hole surrounded by luminous material would look from close-up views. The images must experience extraordinary optical deformations due to the deflection of light rays produced by the strong curvature of the space-time in the vicinity. General relativity allows the calculation of such effects, both on a surrounding accretion disk and on the background star field. This article is an exhaustive and illustrated review of the numerical work on black hole imaging done during the first thirty years of its history.

## Introduction

Black holes are to many the most fascinating objects in space. According to the laws of general relativity they are by themselves invisible. Contrarily to non-collapsed celestial bodies, their surface is neither solid nor gaseous ; it is an immaterial border, the event horizon, beyond which gravity is so strong that nothing can escape, not even light.

Seen in projection on a sky background, the event horizon would have the aspect of a perfectly circular black disk if the black hole is static (Schwarzschild solution) or a slightly distorted one if it is in rotation (Kerr solution). Due to strong gravitational lensing, bare black holes could leave observable imprints on a starry background field in the vicinity of their event horizon, but due to their relatively small sizes and their very large distances from Earth, such imprints are completely out-of-reach of present-day telescopes. Now, in typical astrophysical conditions, a realistic black hole, whatever its size and mass (ranging from stellar to galactic scales), is rarely bare but is dressed in gaseous material. Pulled in a spiral motion, the gas forms a hot accretion disk within which it emits a characteristic spectrum of electromagnetic radiation (Pringle and Rees 1972). Giant black holes, such as those currently lurking at the centers of galaxies, can also be surrounded by a stellar cluster whose orbital dynamics is strongly influenced, whereas stars penetrating into the tidal radius are disrupted (Hills 1975 ; Carter and Luminet 1982). In short, if a black hole remains by itself invisible, it « switches » on in a characteristic way the matter it attracts and distorts the background starry field by gravitational lensing.

Tantalizing progress has been done to indirectly detect black holes through electromagnetic radiation from infalling matter (Pounds et al. 2018) or gravitational waves (Abbott et al. 2016). The Event Horizon Telescope (Doeleman 2008) is an international project whose goal is to take telescopic images at millimeter wavelengths of the nearest massive black hole lurking at the center of our Milky Way galaxy, SgrA* (4.4 million solar masses) and of the supermassive one (6 billion solar masses) in the core of the giant elliptical galaxy M87. The first images at the resolution scale of their apparent event horizons are expected for soon (Doeleman 2017, Goddi et al. 2017).

Well before such an astronomical achievement and as soon as the basics of black holes astrophysics were developed in the 1970's, physicists logically wondered what could look like a black hole surrounded by luminous material — stars or gas. Many educational or artistic representations can be seen in popular science magazines and on the web, in the form of a black sphere floating in the middle of a circular whirlwind of brilliant gas. So striking they are, these images fail to report the astrophysical reality. This one can be correctly described by means of numerical simulations, taking into account the complex distortions that the strong gravitational field prints in space-time and light rays trajectories. The first calculations started more than 40 years ago. Existing reviews on the story of black hole imaging (James et al. 2015, Falcke 2017, Riazuelo 2018, Davelaar et al. 2018, Shaikh 2018) give only references to a few landmarks. The purpose of the present article is to provide an exhaustive and illustrated survey of the work done during the first thirty years of black hole imaging history. As I took an active part into it I'll also give some personal recollections and present unpublished visual material.

## Preliminary steps

Black hole imaging started in 1972 at a Summer school in Les Houches (France). James Bardeen, building on earlier analytical work of Brandon Carter, initiated research on gravitational lensing by spinning black holes (Bardeen 1973). He gave a thorough analysis of light-ray propagation around a Kerr black hole. The Kerr solution for a rotating black hole had been discovered ten years before (Kerr 1963) and since then focused the attention of many searchers in general relativity, because it represented the most general state of equilibrium of an astrophysical black hole (Carter 1971).

The Kerr space-time's metric depends on two parameters : the black hole mass $M$ and its normalized angular momentum $a$. An important difference with usual stars, which are in differential rotation, is that Kerr black holes are rotating with perfect rigidity : all the points on their event horizon move with the same angular velocity. There is however a critical angular momentum $a = M$ (in units where G=c=1) above which the event horizon would « break up » : this limit corresponds to the horizon having a spin velocity equal to the speed of light. For such a black hole, called « extreme », the gravitational field at the event horizon would cancel, because the inward pull of gravity would be compensated by the huge repulsive centrifugal forces.

Bardeen computed how the black hole's rotation affects the shape of the shadow that the event horizon casts on light from a background star field. For a black hole spinning close to the maximum angular momentum, the result is a D-shaped shadow.

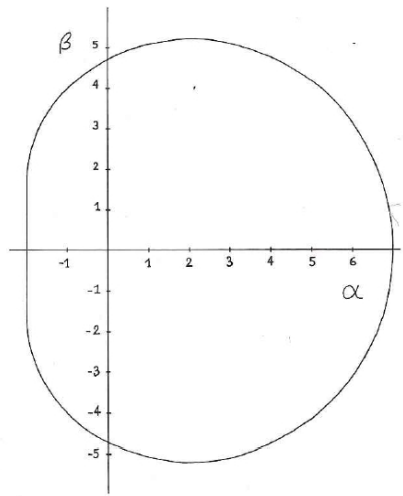

**Figure 1.** Apparent shape of an extreme Kerr black hole as seen by a distant observer in the equatorial plane, if the black hole is in front of a source of illumination with an angular size larger than that of the black hole. The shadow bulges out on the side of the hole moving away from the observer and squeezes inward and flattens on the side moving toward the observer (from Bardeen 1972).

At the time, Chris Cunningham was preparing a PhD thesis at the University of Washington in Seattle under the supervision of Bardeen. He began to calculate the optical appearance of a star in circular orbit in the equatorial plane of an extreme Kerr black hole, taking account of the Doppler effect due to relativistic motion of the star, and pointed out the corresponding amplification of the star's luminosity. He gave formulas but did not produced any image (Cunningham & Bardeen 1972).

One year later the duo published a more complete article (Cunningham & Bardeen 1973) in which, for the first time, a picture was shown of the primary and secondary images of a point source moving in a circular orbit in the equatorial plane of an extreme Kerr black hole. They calculated as functions of time the apparent position and the energy flux of the images seen by distant observers.

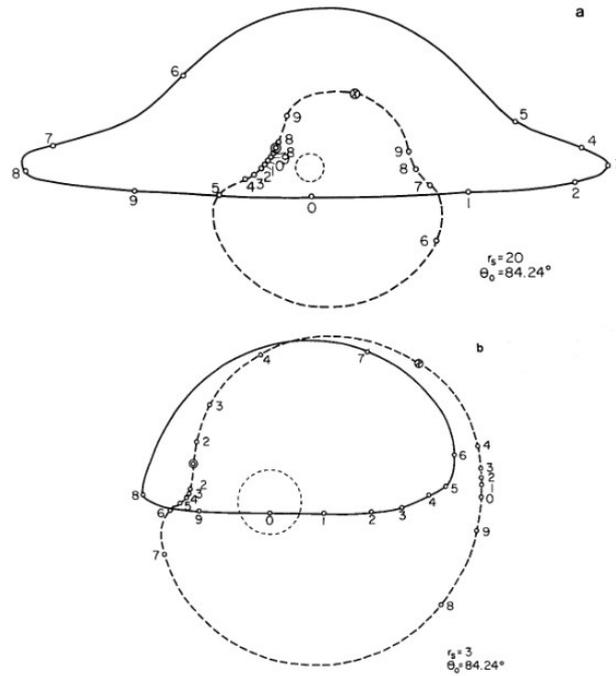

**Figure 2**. Apparent positions of the two brightest images of a point like star in circular orbit as functions of time, for two orbital radii and an observer at a polar angle 84°.024. The small, dashed circle in each plot gives the scale of the plot in units of $GM/c^2$. The direct image moves along the solid line, the secondary image along the dashed line. Ticks mark the positions of the images at 10 equally spaced times (from Cunningham & Bardeen 1973).

In the upper diagram showing the distorted image of a circle of radius 20M, we clearly see that, whatever the observer's inclination angle, the black hole cannot mask any part of the circle behind. We also see that the black hole's spin hardly affects the symmetry of the primary image (although the asymmetry is stronger for the secondary image).

In 1975, Cunningham completed his PhD thesis by calculating the effects of redshifts and focusing on the spectrum of a thin accretion disk around a Kerr black hole. He gave formulas and drew graphics but no image (Cunningham 1975).

The first visualization of the gravitational lensing produced by a non-spinning black hole on a background star field was carried out in 1978 by Leigh Palmer, Maurice Pryce and William Unruh, using an Edwards and Sutherland Vector graphics display at Simon Fraser University (Palmer et al. 1978). They showed a film clip in a number of lectures in that period (including the 9th Texas Symposium on Relativistic Astrophysics), but they did not publish their simulation, so that I can't reproduce any of their images.

## First imaging of a black hole accretion disk

The same year 1978 but quite independently, as a young researcher at Paris-Meudon Observatory specialized in the mathematics of general relativity and following a suggestion from my former PhD advisor Brandon Carter, I wondered what could be the aspect of a Schwarzschild black hole surrounded by a luminous accretion disk.

Accretion disks are expected to form in some double-star systems that emit X-ray radiation (with black holes of a few solar masses) and in the centers of many galaxies (with black holes whose mass adds up to between one million and several billion solar masses). Their close-up images must experience extraordinary optical deformations, due to the deflection of light rays produced by the strong curvature of the space-time in the vicinity of the black hole. General relativity and a computer allow to calculate such an effect.

Then I undertook to compute the bolometric appearance (i.e. including the contributions at all electromagnetic wavelengths) of a thin accretion disk gravitationally lensed by a non-spinning black hole, as seen from far away, but close enough to resolve the image. I considered a Schwarzschild black hole and a physical model of a thin disk of gas viewed from the side, either by a distant observer or a photographic plate.

Throughout my researchers' career I always considered that, before writing a computer program with full equations and running the machine, pure geometrical reasoning is of a great help to get a preliminary idea of the result. Otherwise, what guarantee have you that your program had no mistake ?

I did this kind of geometrical reasoning to guess what I should obtain for a black hole's accretion disk appearance. In an ordinary situation, meaning in Euclidean space, the curvature is weak. This is the case for the solar system when one observes the planet Saturn surrounded by its magnificent rings, with a viewpoint situated slightly above the plane. Of course, some part of the rings is hidden behind the planet, but one can mentally reconstruct their elliptic outlines quite easily. Around a black hole, everything behaves differently because of the optical deformations due to the space-time curvature, as depicted in the diagram below - published only years later in my popular book on black holes (Luminet 1992).

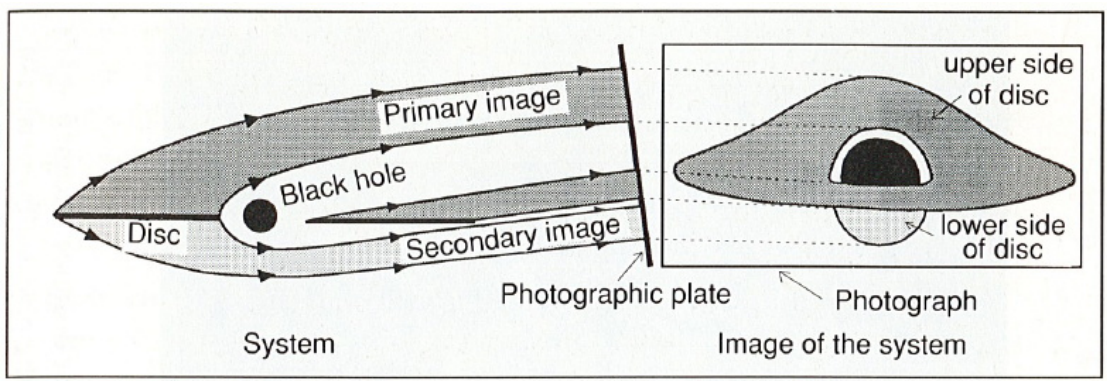

**Figure 3.** Optical distortions near a black hole. We imagine a black hole surrounded by a bright disk. The system is observed from a great distance at an angle of 10° with respect to the plane of the disk. The light rays are received by a photographic plate. Because of the curvature of space-time in the neighborhood of the black hole, the image of the system is very different from the ellipses which would be observed if an ordinary star or a planet replaced the black hole. The light emitted from the upper side of the disk forms a direct image and is considerably distorted so that it is completely visible. The lower side of the disk is also visible as a secondary image caused by highly curved light rays (from Luminet 1992).

Strikingly, we can see the top of the disk in its totality, whatever the angle from which we view it may be. The back part of the disk is not hidden by the black hole, since the images that come from it are to some extent enhanced by the curvature, and reach the distant observer. Much more astonishing, one also sees a part of the bottom of the gaseous disk. In fact, the light rays which normally propagate downwards, in a direction opposite to that of the observer, climb back to the top and give a « secondary image », a highly deformed picture of the bottom of the disk. We also check that no radiation can come from the region between the black hole's event horizon and the inner edge of the disk, because the properties of the Schwarzschild space-time forbid the accretion disk from touching the surface of the black hole. The circular orbits of the gas in the disk can be maintained only down to a critical distance of three times the Schwarzschild radius (*i.e.* 6M). Below this the disk is unstable; the gas particles plunge directly towards the black hole without having enough time to emit electromagnetic radiation.

Then I could start to calculate the things numerically. I used the IBM 7040 mainframe of Paris-Meudon Observatory, an early transistor computer with punch card inputs. The machine generated isolines that were directly translatable as smooth curves using the drawing software available at the time. The first step was to integrate the equations of light ray trajectories in Schwarzschild space-time and draw the isoradial curves (*i.e.* at constant radial distance from the black hole) of a thin disk around the black hole, as they would be seen by an observer above the disk's plane (Figure 4).

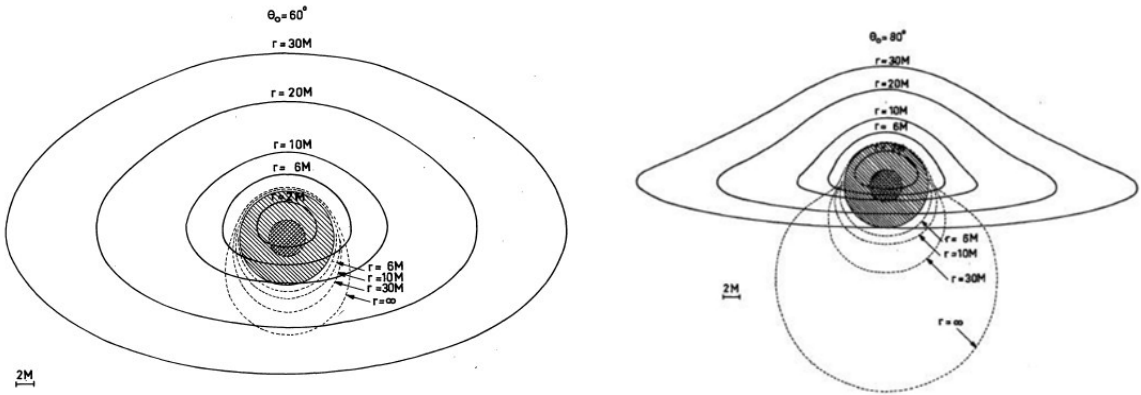

**Figure 4.** Isoradial curves representing rays emitted at constant radius from the hole, as seen by an observer at 30° (Left) and 10° (Right) above the disk's plane. Full lines : direct images; dashed lines : secondary images.

In fact the gravitational lensing generates an infinity of images of the disk, because the light rays can travel any number of times around the black hole before escaping from its gravitational field and being observed by a distant astronomer. The primary image shows the upper side, the secondary image shows the lower side, the third image shows the upper side again, and so on. However, higher order images are not optically interesting because they are stuck to the edge of the central black spot, the latter representing the « shadow » of the actual black hole, namely the apparent size of its event horizon expanded by a factor $3\sqrt{3}/2 \sim 2.6$.

By comparison with the image calculated by Cunningham and Bardeen for a circular ring at distance 20M around an extreme Kerr black hole (Fig. 2), we check that the spin of the black hole hardly affects the shape of the primary image.

The next step was to take into account the physical properties of the gaseous disk : rotation, temperature and emissivity. When the flow of matter into a black hole is not too great, it forms a very thin accretion disk. In that case, the proper luminosity of the disk can be accurately calculated, according to models first described in 1973 (Shakura & Sunyaev 1973) and 1974 in their relativistic version (Page and Thorne 1974): the intensity of radiation emitted at any given point of the disk only depends on its temperature, and the latter only depends on the radial distance to the black hole. Therefore the intrinsic brightness of the disk cannot be uniform. The maximum luminosity comes from the inner regions close to the event horizon, because it is there that the gas is the hottest.

In addition, the apparent luminosity of the disk is still very different from its intrinsic one: the radiation picked up at a great distance is frequency- and intensity-shifted with respect to the emitted one. There are two sorts of shift effects. There is the Einstein effect, in which the gravitational field lowers the frequency and decreases the intensity, and there is the Doppler effect, where the displacement of the source with respect to the observer causes amplification as the source approaches and attenuation as the source retreats. The combination of the two effects is not trivial.  I calculated it at each point of the photographic plate. In our case the Doppler effect dominates, and it is caused by the disk rotating around the black hole. The regions of the disk closest to the black hole rotate at a velocity approaching that of light, so that the Doppler shift is considerable and drastically modifies the image as seen by a distant observer. The sense of the rotation of the disk is such that matter recedes from the observer on the right-hand side of the photograph and approaches on the left-hand side. As the matter recedes, the Doppler deceleration is added to the gravitational deceleration, explaining the very strong attenuation on the right-hand side. In contrast, on the left-hand side the two effects tend to cancel each other out, so the image more or less retains its intrinsic intensity. Figures 6-7 depict the isophotes - i.e. the curves of constant apparent luminosity - as seen by an observer at respectively 30° and 10° above the disk's plane.

The final black and white "photographic" image was obtained from this pattern. Lacking of an appropriate drawing software, I had to create it by hand. Using numerical data from the computer, I drew directly on negative paper with pen and Indian ink, placing dots more densely where the simulation showed more light (a few thousands dots for the full plate). Next, I took the negative of my negative to get the positive, the black points becoming white and the white background becoming black. The result converged into the pleasantly organic, asymmetrical form reproduced in Figure 8, both visually engaging and scientifically revealing.

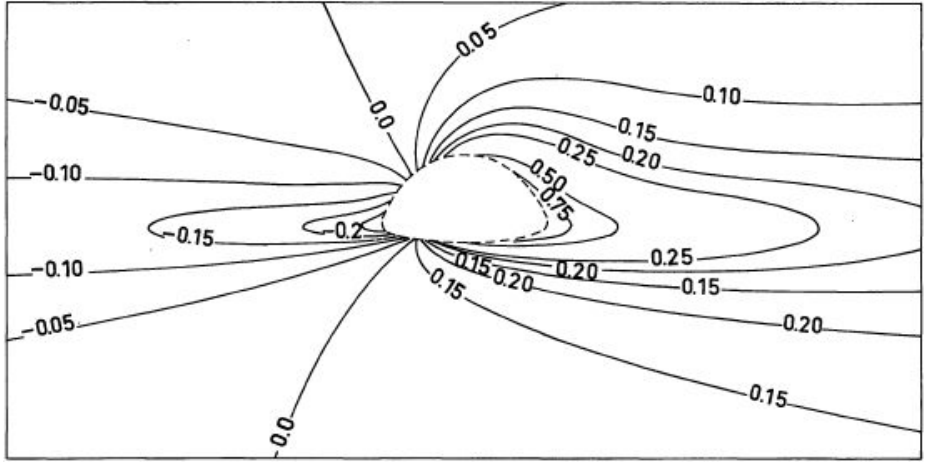

**Figure 5.** Curves of constant spectral shift, as seen by an observer at 10° above the plane of the disk. Positive values of the spectral shift contours correspond to redshifts, negative values to blueshifts (from Luminet 1979).

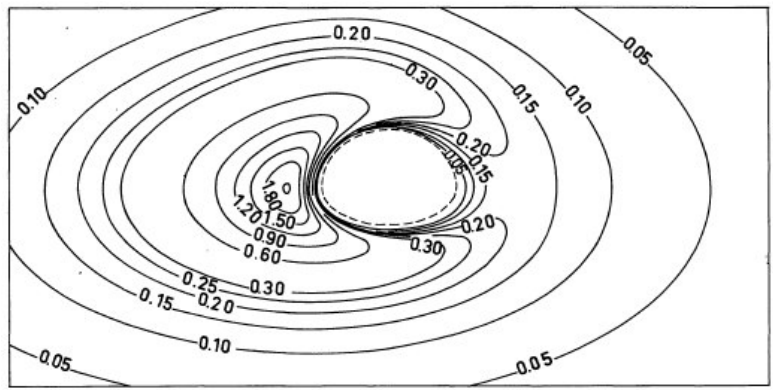

**Figure 6.** Curves of constant apparent flux in units of the maximum intrinsic flux, as seen by an observer at 30° above the disk's plane. Dashed line : the apparent inner edge of the disk at r=6M, where flux is zero (from Luminet 1979).

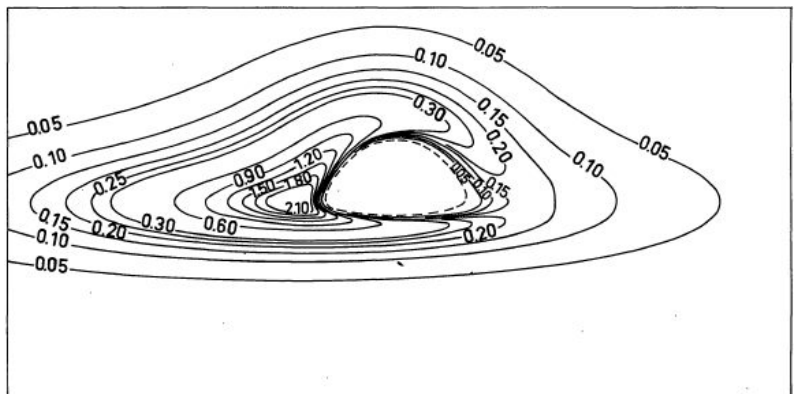

**Figure 7.** Curves of constant apparent flux in units of the maximum intrinsic flux, as seen by an observer at 10° above the disk's plane. The luminosity is maximum in the regions where high intrinsic luminosity combines with high blueshift (from Luminet 1979).

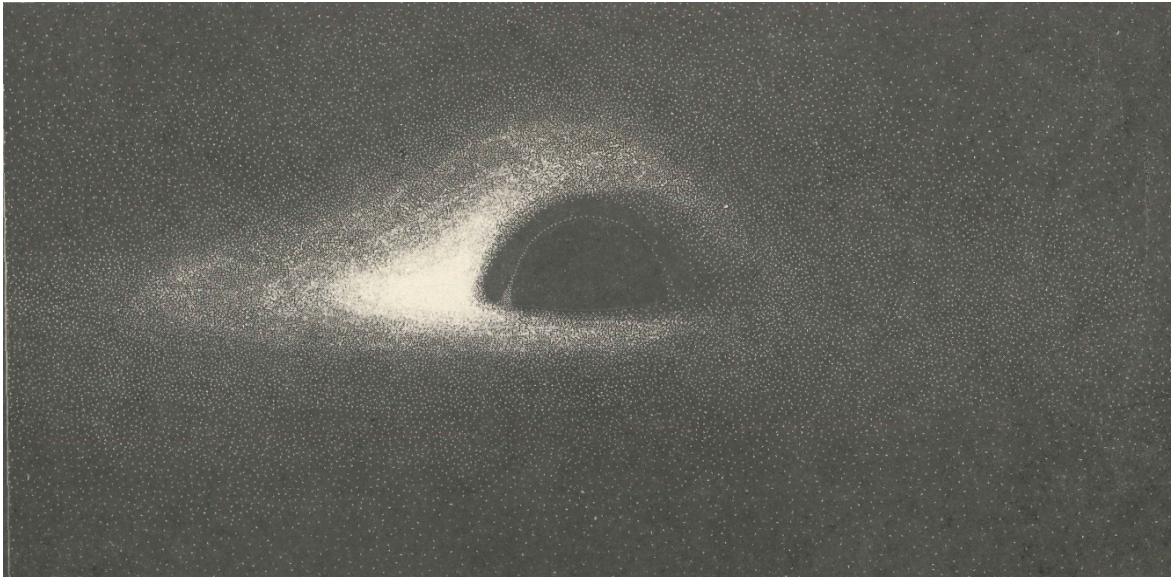

**Figure 8.** Image of a Spherical Black Hole with Thin Accretion Disk. As expected, the main characteristic of the image is the strong asymmetry of the disk's brightness, so that one side is far brighter and the other is far dimmer. Another one is that, although the upper side of the disk is completely visible, only a small part of the lower side is observable. This is due to the fact that in a realistic situation, the gaseous disk is opaque; therefore it absorbs the light rays that it intercepts. It follows that a major part of the secondary image is occulted by the primary image, its visible part being stuck around the edge of the black hole, like a gleaming halo (from Luminet 1979).

To comment in a non-scientific manner this first theoretical glimpse of the shadow of a black hole, no caption could fit better than these verses by the French poet Gérard de Nerval, written as soon as in 1854 (Nerval 1854) :

*In seeking the eye of God, I saw nought but an orbit*
 *Vast, black, and bottomless, from which the night which there lives*
*Shines on the world and continually thickens*

*A strange rainbow surrounds this somber well,*
*Threshold of the ancient chaos whose offspring is shadow,*
*A spiral engulfing Worlds and Days !*

The picture was first published in November 1978 a popular article on black holes for a French magazine (Carter & Luminet 1978), and the complete work with technical details a few months later in a peered-reviewed European journal (Luminet 1979). A funny anecdote is that many readers who saw for the first time this simulated picture of a perfectly *non-luminous* star believed that the author used the name *Luminet* as a pseudonym. Indeed I enjoyed the pun !

## An anticipated image of Sagittarius A* ?

The 1978 calculation was independent of the mass of the black hole and of the flow of gas swallowed, on the condition that the latter remains moderate. In other situations, found for instance in binary X-ray sources harboring stellar mass black holes or in active galactic nuclei with supermassive black holes accreting near or above the Eddington limit, the structure of the accretion disk may be thick, take the form of a torus and so on (Abramowicz et al. 1978). The picture of figure 8 could therefore correctly describe only relatively weak sources, like giant black holes lying at the center of non- active galaxies and sucking in the interstellar gas at a mild rate.

At the time of the calculation, it had just been suggested that the giant elliptical galaxy M87 could harbor such a giant black hole with a few billion solar masses (Sargent et al. 1978) and that the newly discovered compact radiosource Sagittarius A* (SgrA*) at the center of our own galaxy, the Milky Way, could be connected to a massive black hole of a few million solar masses (Balick & brown 1974). The observational support was still very weak, but since then we have convincing indications that most of non-active galactic nuclei harbor « dormant » massive black holes (Kormendy & Ho 2001) surrounded by thin gaseous disks accreting at a low rate. For instance, the best estimate of the accretion rate onto SgrA* is provided by radio polarization measurements and is constrained to vary in the range $10^{-9}M_S/yr$ — $10^{-7}M_S/yr$ on scales of hundreds to thousands Schwarzschild radii (Marrone et al. 2007). Thus SgrA* is an ideal target to check the accuracy of Figure 8, although I did not suspect this : the idea to image in practice the shadow of SgrA* using VLBI would not be presented before 2000 (Falcke et al. 2000).

Compared to the rough simulation of 1979, a realistic image reconstruction of the galactic black hole SgrA* involves a number of additional restrictions. For instance, as already mentioned, my computerized black hole picture included the contributions of all electromagnetic wavelengths, whereas any observational device works only in a restricted range of frequencies. The theory of thin accretion disks tells us that the wavelengths at which the intrinsic luminous flux is maximal depends on the mass of the black hole. For a stellar mass black hole it should be in the X-ray band, but for a several millions solar masses black hole, the major part of emission would be in the millimeter radio band.

Now go back to figure 6 showing curves of constant apparent flux seen by an observer at 30° above the disk's plane. Assume that the black hole is about 4 millions solar masses, as suggested from recent measurements of SgrA* (Chatzopoulos et al. 2015) and assume that we have a VLBI network of radiotelescopes at the limit of sensitivity to detect such a radiation. Due to the various shift effects described above, only the part of the image with an apparent flux greater than 1.0 (in units of the maximum intrinsic flux) will be detectable. The remaining image would be the bright crescent shown in Figure 9. It is interesting to point out the similitude of this rough result with some recent image reconstructions of Sgr A* performed in the framework of the Event Horizon Telescope program, through state-of-art multi-dimensional general relativistic magneto-hydrodynamic simulations (Johnson et al. 2017).

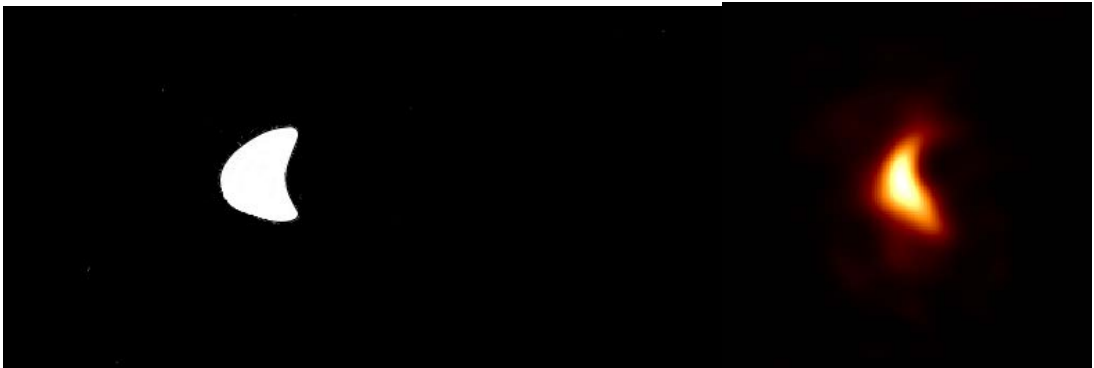

**Figure 9.** Left : Remaining image of Figure 6 when the regions with an apparent flux lower than 1.0 have ben masked. Right : Image reconstruction of SgrA* Synthetic that could be recovered with an array of 8 radiotelescopes and averaging 8 epochs, using general relativistic magneto-hydrodynamic simulations, for a similar angle of view (from Johnson et al. 2017).

## Colors bloomed first in Japan

Go back to the thread of history. Black hole imaging remained a backwater of physics research until a decade later. In 1987, two educational papers on gravitational lensing by black holes appeared in a journal for physics teachers and students. Schastok et al (1987) discussed the effects of light deflection on the stellar sky as seen from the vicinity of a rotating black hole and showed some light rays circling an extreme Kerr black hole, but they did not produce a realistic picture. Ohanian (1987) treated the case when the light can circle around a Schwarzschild black hole one or several times, giving rise to a sequence of images. Apparently ignoring my 1979 article, he provided the same formulas for the deflection angles and light intensities in terms of elliptic integrals, and he did not show any visualization.

In 1988 I travelled to Japan to attend an international conference in Tokyo. Two young Japanese searchers came to me and, to my surprise, they wanted to take a photograph of all three together. Indeed, Jun Fukue and Takushi Yokoyama had just extended my work by adding (false) colors to the black hole picture (Fukue & Yokohama 1988). This was a real improvement, since my monochromatic (because bolometric) image did not take into account the real changes of apparent wavelengths in the accretion disk, shifted from blue on one side to red on the other, and so on. Here are some of the pictures they produced. The Japanese boys told me that in their manuscript, they initially used the term "Lady" for the accretion disk, but the editors replaced this by the "accretion disk", in accordance with the general style of the *Publications of the Astronomical Society of Japan*. It was well before the feminization of language that prevails today in every branch of human activity, including astrophysics!

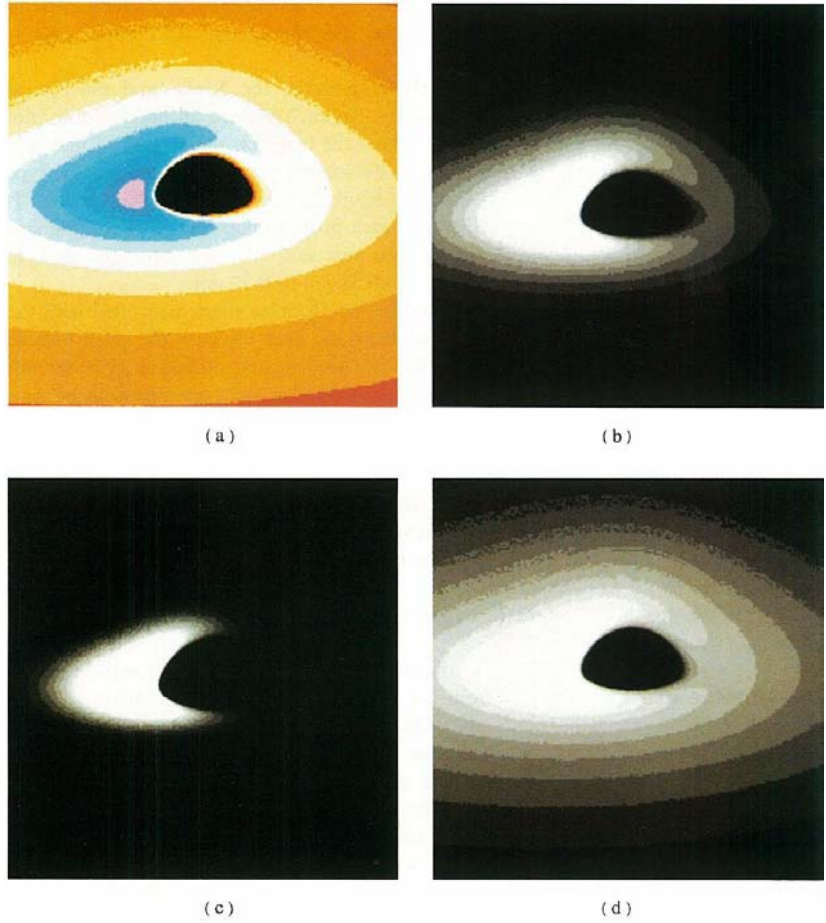

**Figure 10.** Four types of photographs of a disk with $T_{max} = 10^7$ K rotating a stellar-mass Schwarzschild black hole. The photographer's distance from the center is $10^4$ times the Schwarzschild radius, and is inclination angle is 20°. (a) The distribution of the temperature in false colors, from orange ($T_{obs} < 3.10^6$ K) through yellow ($4.10^6$ K $< T_{obs} < 5.10^6$ K) and white ($6.10^6$ K $< T_{obs} < 7.10^6$ K) to violet ($T_{obs} > 10^7$ K). (b) The bolometric photograph.(c) The X-ray photograph in the 2-30-keV band. (d) The optical photograph.

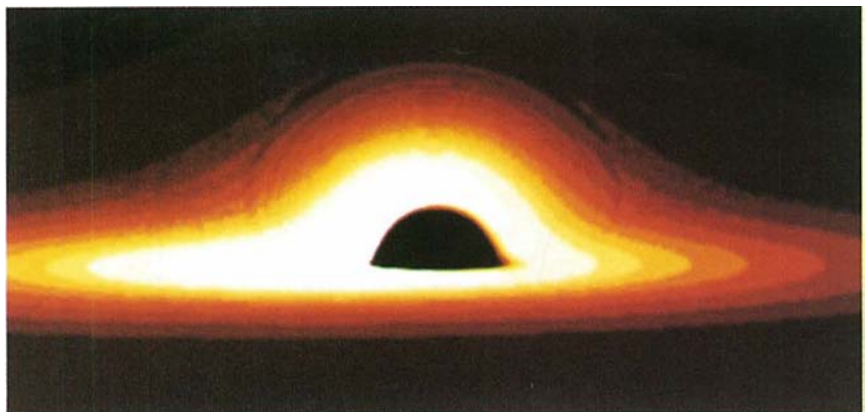

**Figure 11.** The color photograph of a disk located at the center of active galaxies

# First Flight into a Black Hole

In 1989-1990, while I spent one year as a research visitor at the University of California, Berkeley, my former collaborator at Paris-Meudon Observatory, Jean-Alain Marck, both an expert in general relativity and computer programming, started to extend my simulation of 1979. The fast improvement of computers and visualization software (he used a DEC-VAX 8600 machine) allowed him to add colors and motions. To reduce the computing time, Marck developed a new method for calculating the geodesics in Schwarzschild space-time, published only several years later (Marck 1996). In a first step (Marck 1989), he calculated static images of an accretion disk around a Schwarzschild black hole according to various angles of view, see Figure 12.

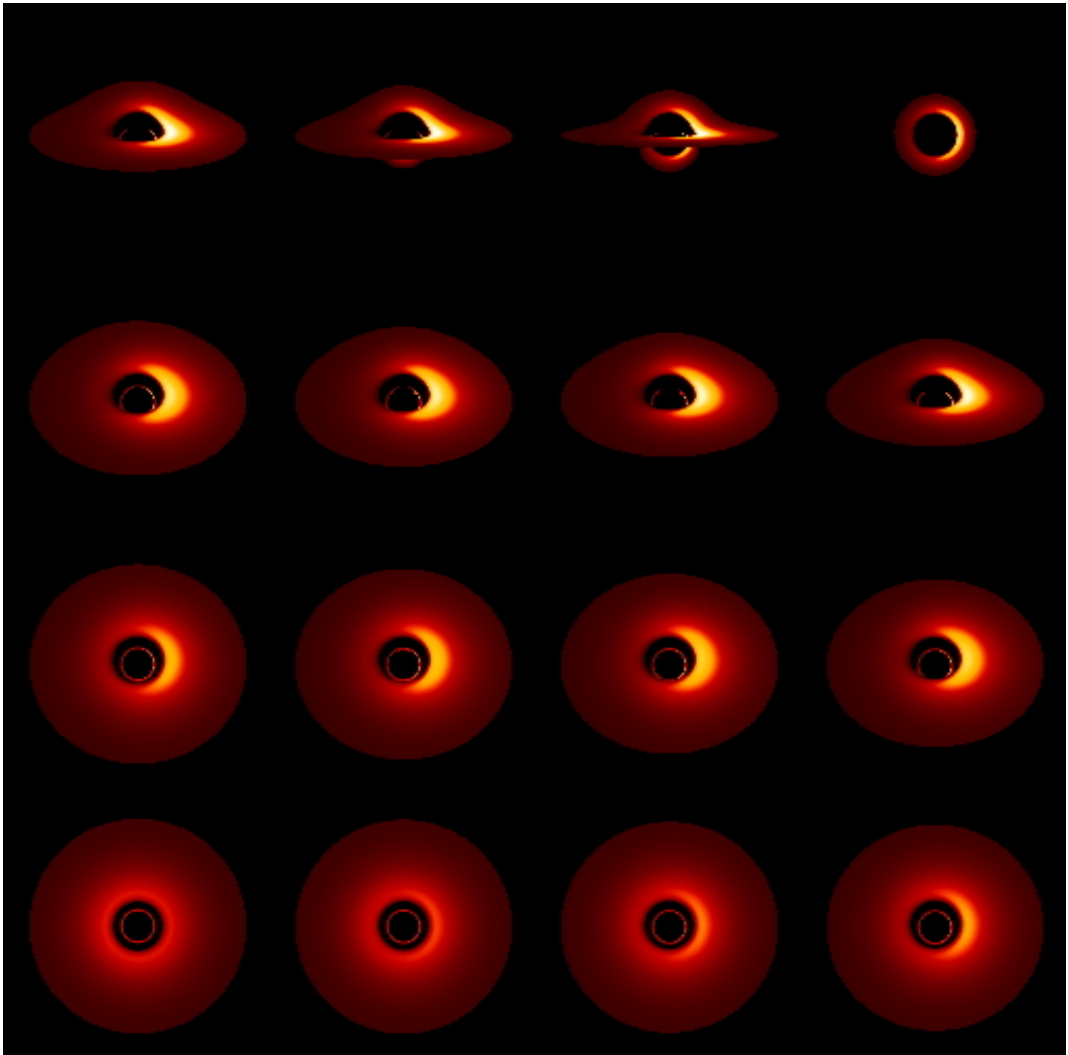

**Figure 12.** Colored images of a black hole accretion disk for various angles of view by J.-A. Marck, 1989 (unpublished).

In 1991, when I went back to Paris Observatory, I started the project for the French-German TV channel Arte of a full-length, pedagogical movie about general relativity (Delesalle et al. 1994). As the final sequence dealt with black holes, I asked Marck to introduce motion of the observer with the camera moving around close to the disk, as well as to include higher-order lensed images and background stellar skies in order to make the pictures as realistic as possible. The resulting animation for an observer moving along an elliptic trajectory around a Schwarzschild black hole and crossing several times the plane of a thin accretion disk (Figure 13) is now available on the web (Marck 1991). Compared to the static, black-and-white simulation of 1979, the snapshot reproduced in Figure 14 shows spectacular improvements.

Eventually, for the needs of a videocassette shown in various international conferences and workshops, Marck calculated a new series of images as seen by an observer plunging into the event horizon of a Schwarzschild black hole along a parabolic trajectory (Marck 1994). A few snapshots were reproduced later in a special issue of a French popular magazine devoted to black holes (Marck & Luminet 1997), see Figure 15.

The black hole visualizations obtained by Jean-Alain Marck not only were a very significant improvement of all previous work, but they would remain unsurpassed for about twenty years, both scientifically and aesthetically. As a striking illustration, Figure 16 compares the view of an accretion disk calculated by Marck for an observer in the equatorial plane of a Schwarzschild black hole, including all the shift effects, a truly physical model of the accretion disk and effects of light diffusion, and the famous view designed in 2014 for the science-fiction movie *Interstellar,* calculated with similar conditions (except the fact that the black hole was of the Kerr type, but as already pointed out the rotation does not affect significantly the asymmetry of the image). The latter was obtained by a team of 200 graphic animation experts who used a general relativistic programming code provided by their scientific advisor Kip Thorne (James et al. 2015, Thorne 2014a). At the time it was presented by some medias as a « simulation of unprecedented accuracy » (Rogers 2014).

As I already commented in more details in previous papers (Luminet 2015, Luminet 2018), both images correctly describe the primary and secondary images created by the gravitational field, but the second one, used for *Interstellar* and despite the advice of Kip Thorne (Thorne 2014b), left out the shift effects and the physical structure of the accretion disk on the pretext that a highly asymmetric image of the disk would be harder for a mass audience to grasp. However it is precisely this strong asymmetry of apparent luminosity that is the main signature of the black hole, the only celestial object able to give the internal regions of an accretion disk a speed of rotation close to the speed of light and to induce a very strong shift effect. In short *Interstellar's* image was certainly impressive but was not really what a black hole would look like; Marcks' image was much closer from astrophysical reality, in addition to be, in my opinion, aesthetically even more appealing.

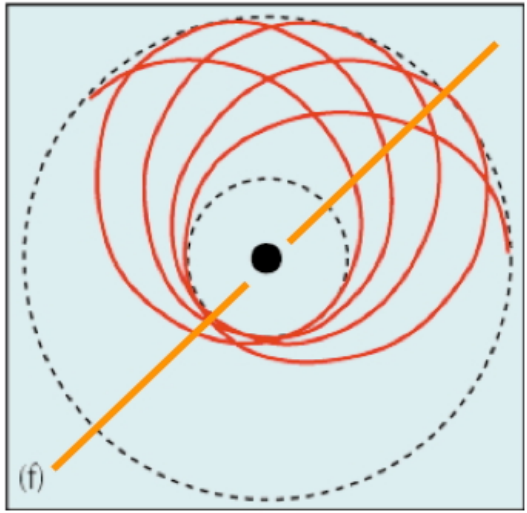

**Figure 13.** Elliptic trajectory around a Schwarzschild black hole followed by the observer in the movie by Delesalle et al. (1994). The trajectory undergoes a strong general relativistic precession of its periastron, and the observer crosses several times the plane of the accretion disk.

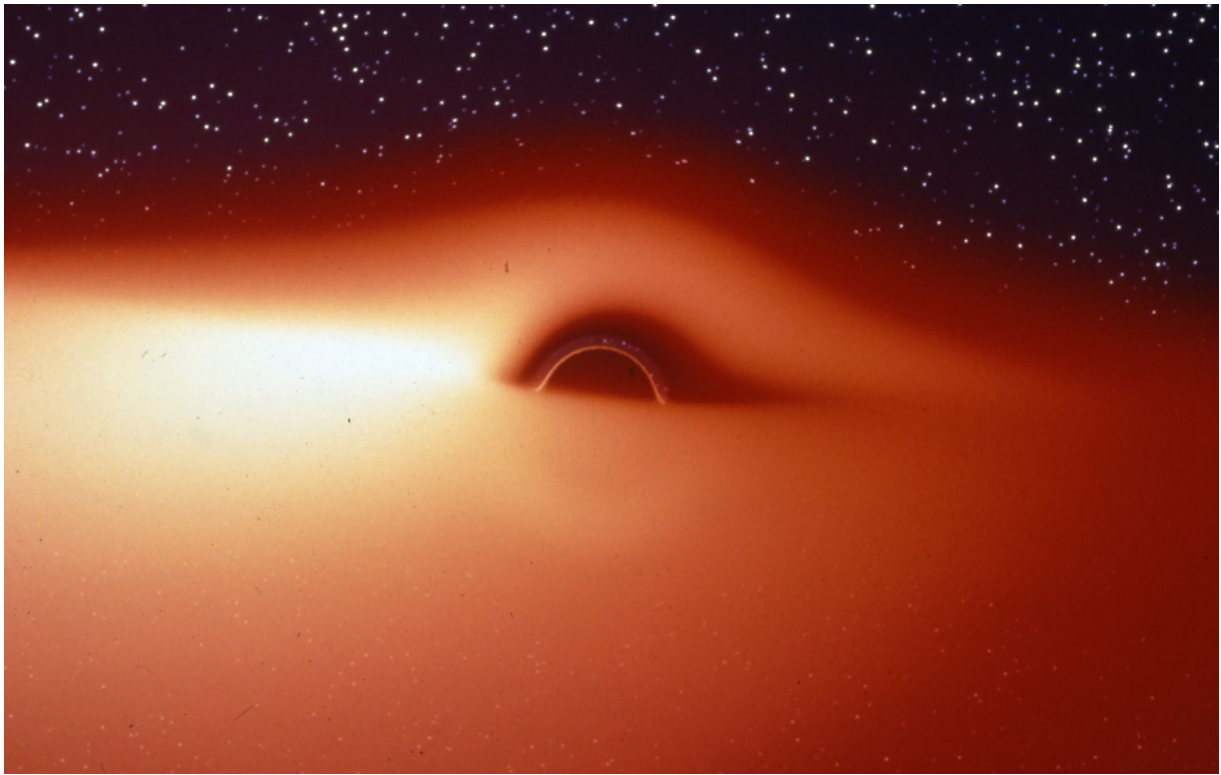

**Figure 14.** Colored image of a black hole accretion disk as seen by a moving observer at 7° above the disk's plane. The observer uses a camera equipped with filters to convert into optical radiation the emitted electromagnetic radiation. The arbitrary coloring encodes the apparent luminosity of the disk, the brightest and warmest parts being colored yellow, the colder parts red. The transparency of the disk was enhanced in order to show the secondary image through the primary, as well as some background stars. Compared with figure 8 there are additional distortions and asymmetries due to the Doppler effect induced by the motion of the observer himself. As a result the region of maximum luminosity has no more the shape of a crescent (from Marck 1991)

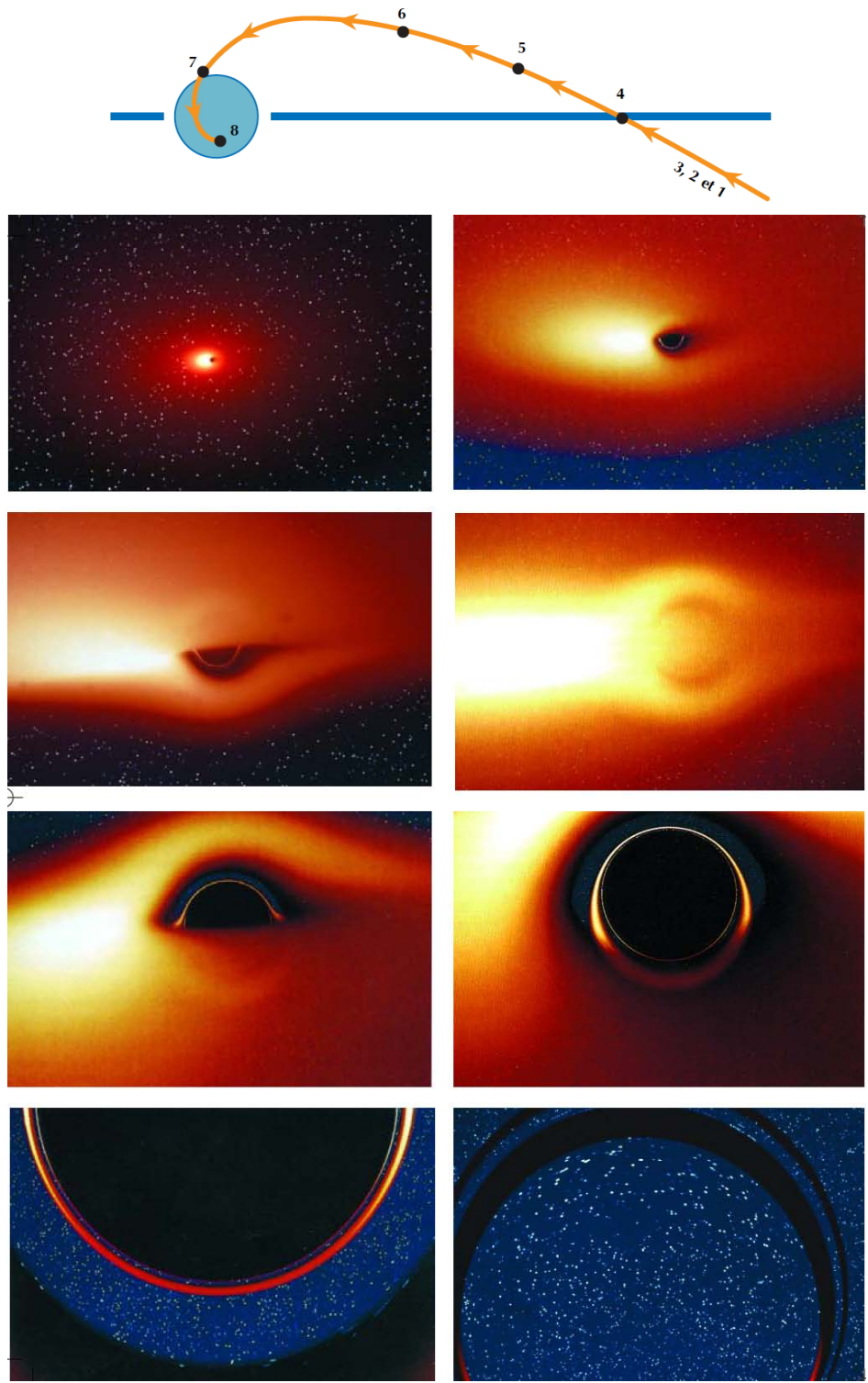

**Figure 15.** This series of snapshots are taken by a traveller at positions 1 to 8 of the parabolic trajectory plotted in the top part. Initially he is located under the plane of the disc, at a distance 1200 M from the center. He crosses the accretion disk at 39 M (4) and is very close to the horizon at point 7. Then his speed approaches that of light and the image distortions become quite considerable. He takes his last shot (8) inside the black hole at 0.7 M from the center, having rotated 180 degrees to look through the rear window and see one last time several strongly distorted images of the accretion disk and the background stellar sky (from Marck & Luminet 1997).

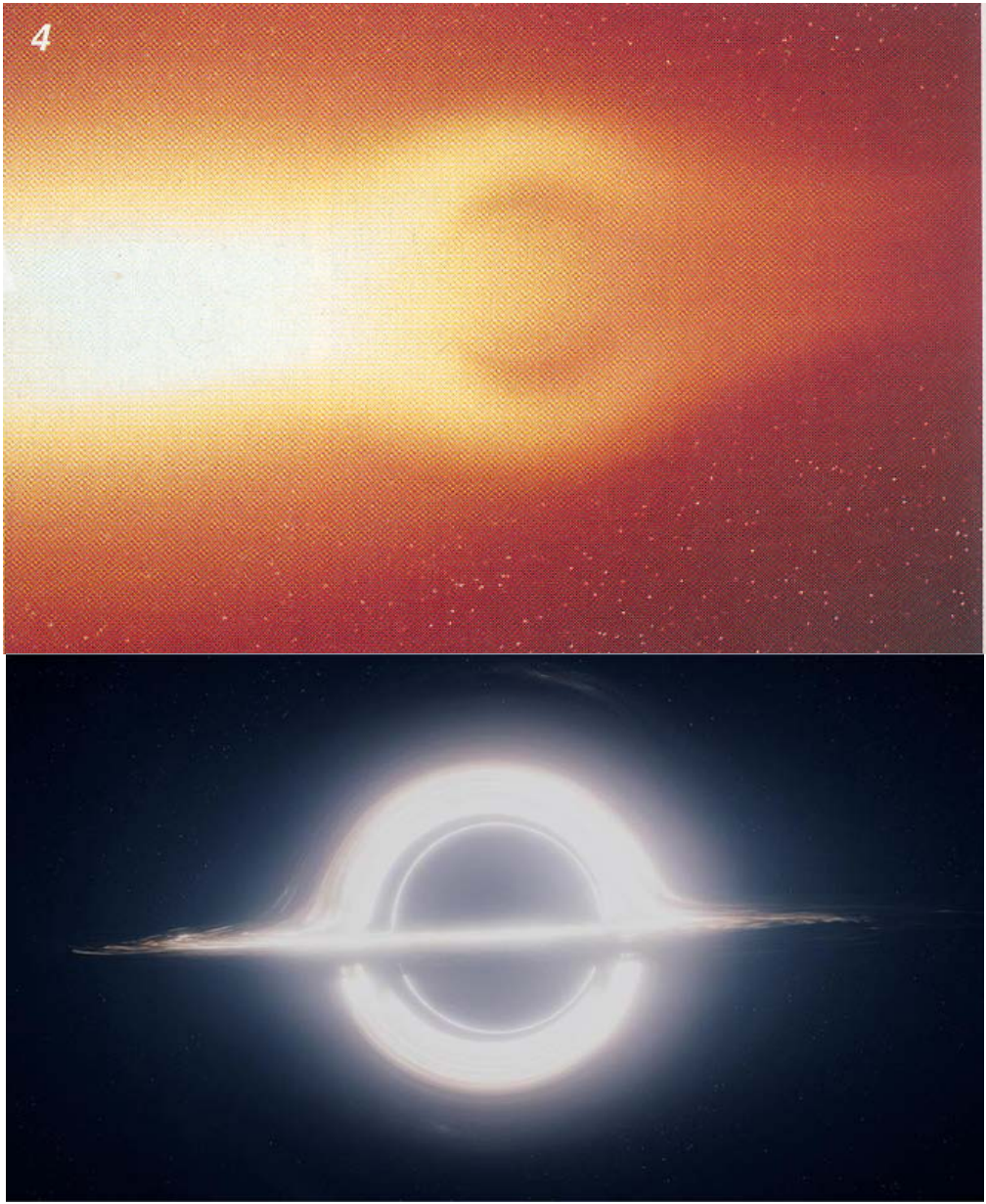

**Figure 16.** Top : Marck's simulation of a black hole accretion disk as seen in the equatorial plane, 1994. Bottom : the same by James et al. for the movie *Interstellar*, 2014.

## Generalizations to Kerr Black Holes

Unfortunately Marck's simulations of black hole accretion disks remained mostly ignored from the professional community, due to the fact that they were not published in peer-reviewed journals and, after their author prematurely died in May 2000, nobody could find the trace of his computer program…

Then, unaware of Marck's results, several researchers of the 1990's were involved in the program of calculating black hole gravitational lensing effects in various situations. Stuckey (1993) studied photon trajectories which circle a static black hole one or two times and terminate at their emission points (« boomerang photons »), producing a sequence of ring-shaped mirror images. Nemiroff (1993) described the visual distortion effects to an observer traveling around and descending to the surface of a neutron star and a black hole, discussing multiple imaging, red- and blue-shifting, the photon sphere and multiple Einstein rings. He displayed computer-generated illustrations highlighting the distortion effects on a background stellar field but no accretion disk, and made a short movie now available on the internet (Nemiroff 2018), two snapshots of which are shown in figure 17.

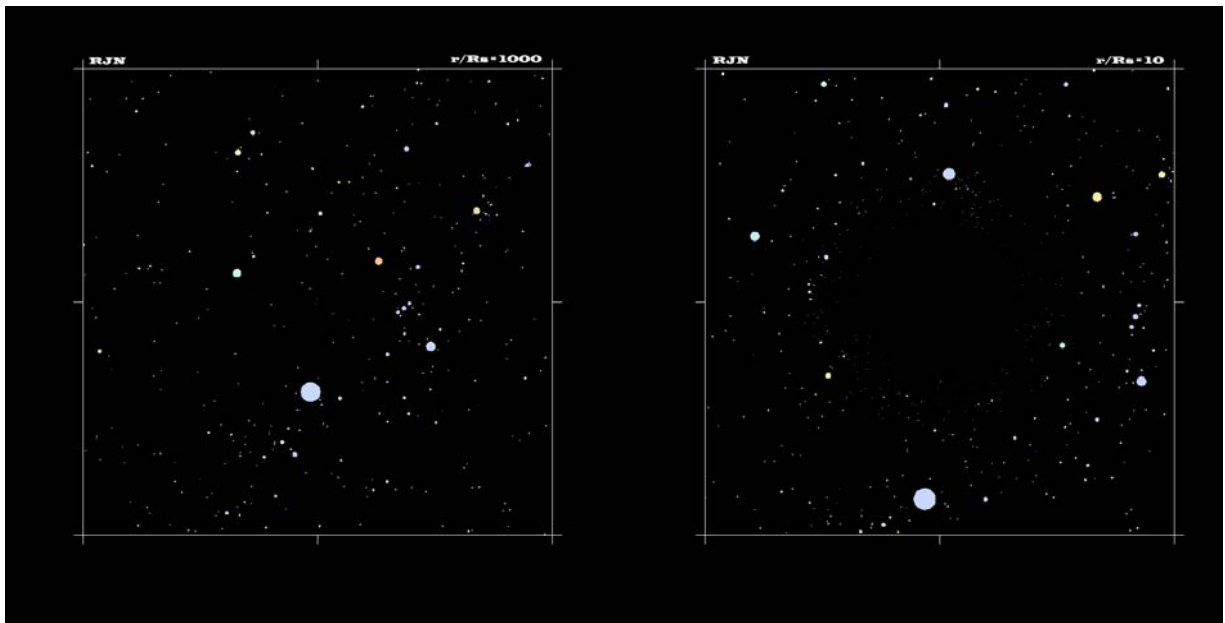

**Figure 17.** Trip to a black hole by Robert Nemiroff, 1993.

The first simulations of the shape of accretion disks around Kerr black holes were performed by Viergutz (1993). He treated slightly thick disks and produced colored contours, including the disk's secondary image which wraps under the black hole (figure 18). The result is a colored generalization of Figure 2 by Cunningham and Bardeen (1973).

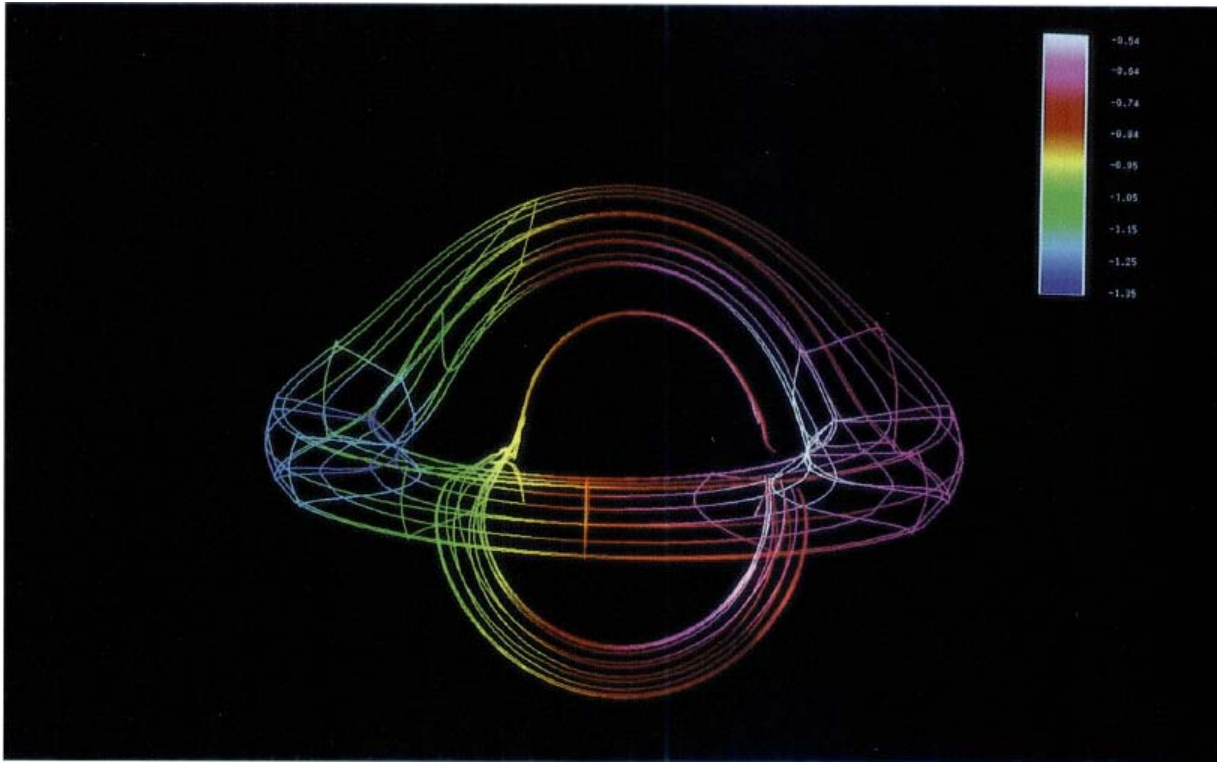

**Figure 18.** Primary and secondary images of a simple accretion disk model around a Kerr black hole, seen by a faraway observer. Colors indicate combined gravitational and Doppler shifts (from Viergutz 1993).

More elaborate views of a geometrically thin and optically thick accretion disk around a Kerr black hole were obtained by Fanton et al. (1997). They developed a new program of ray tracing in Kerr metric, and added false colors to encode the degree of spectral shift and temperature maps (figure 19). Zhang et al. (2002) used the same code to produce black-and-white images of standard thin accretion disks around black holes with different spins, viewing angles and energy bands (figure 20).

Bromley et al. (1997) calculated integrated line profiles from a geometrically thin disk about a Schwarzschild and an extreme Kerr black hole, in order to get an observational signature of the frame-dragging effect (Figure 21).

In 1998 Andrew Hamilton started to develop for a student project at the University of Colorado a "Black Hole Flight Simulator", with film clips that have been shown at planetariums, also available on the Internet. The first depictions were very schematic, but the website was constantly implemented. It now offers journeys into a Schwarzschild or a Reissner-Nordström (i.e. electrically charged) black hole with effects of gravitational lensing on a stellar background field, as well as animated visualizations of magneto-hydrodynamic simulations of a disk and jet around a non-rotating black hole (Hamilton 2018).

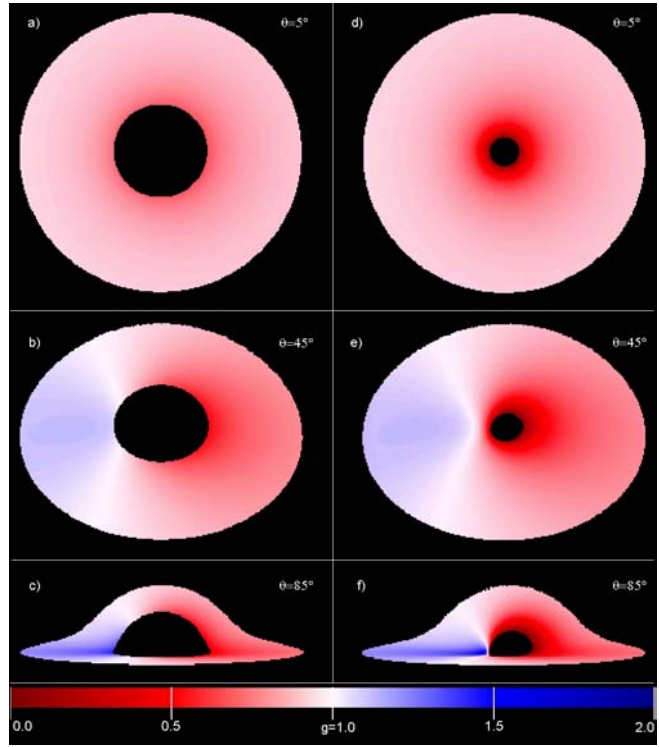

**Figure 19.** False color contour maps showing how the monochromatic radiation emitted by a Keplerian accretion disk would be seen at infinity for various values of the inclination angle to the plane of the disk (top to bottom : 5°, 45°, 85°). The left column refers to a non-rotating black hole, the right one to a rapidly rotating black hole with a=0.998 M. The white zones stand for the regions with zero redshift. Left-hand side of the disk is approaching the observer and blueshifted (from Fanton et al. 1997).

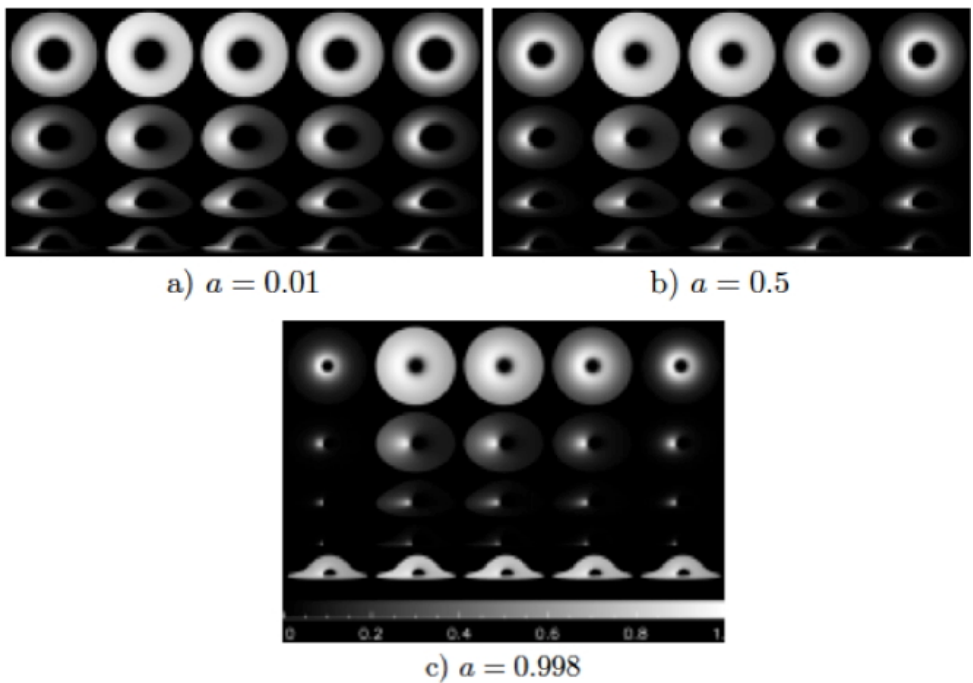

**Figure 20.** Disk images of accretion disks extending up to 20 Schwarzschild radii for different spins of Kerr black holes, viewed in different energy ranges and inclination angles (from Zhang et al. 2002).

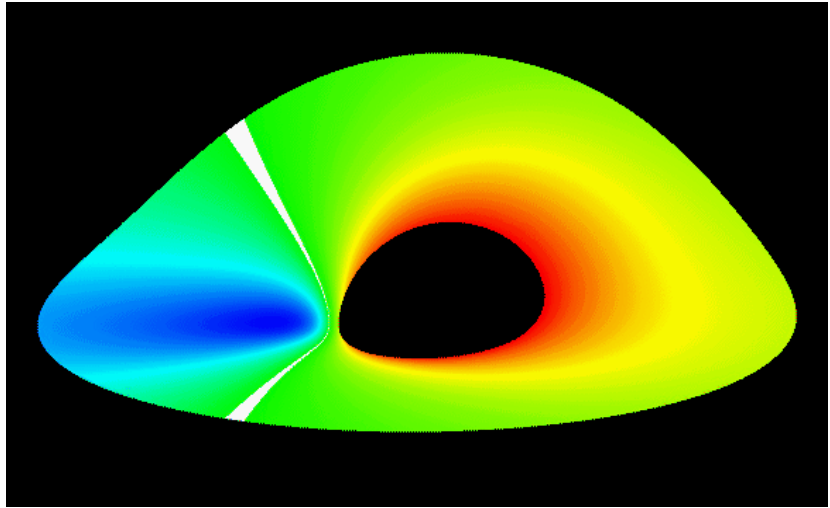

**Figure 21.** Image of a geometrically thin disk around an extreme Kerr (maximally rotating) black hole seen at an inclination of 75°. The inner and outer radii of the Keplerian (circularly rotating) disk are at 1.24 M and 6 M. The colors encode the apparent light frequency, the white strip divides redshifted and blueshifted regions. The asymmetric appearance of the inner disk edge results from the frame-dragging effect of black hole rotation (from Bromley et al. 1997).

## From Idea to Reality

A turning point in the history of black hole imaging came when the possibility of viewing in practice the shadow of SgrA* with VLBI radio astronomy techniques was first discussed (Falcke et al. 2000, Doeleman et al. 2001). Falcke et al. (who did not quote my pioneering article) developed a general relativistic ray-tracing code that allowed them to simulate observed images of Sgr A* for various combinations of black hole spin, inclination angle, and morphology of the emission region directly surrounding the black hole (figure 22).

Bromley et al. (2001) provided maps of the polarized emission of a Keplerian disk to illustrate how the images of polarized intensity from the vicinity of SgrA* would appear in future VLBI observations (Figure 23).

Indeed, in parallel with but rather independently from the theoretical simulations reviewed here, the work to image SgrA* by VLBI experiments had begun also back in the 1970's, after the discovery of the compact radio source Sgr A* at the center of the Milky Way and its identification as the likely emission of gas falling onto a supermassive black hole (Balick and Brown 1974). And as soon as it was realized that the shadow of SgrA* could really be photographed in the forthcoming years, the program of imaging black holes with or without accretion disks and/or stellar background field developed at a much accelerated rate. Several dozens of papers with more or less elaborate visualizations bloomed out, but they do not enter into the purpose of the present review.

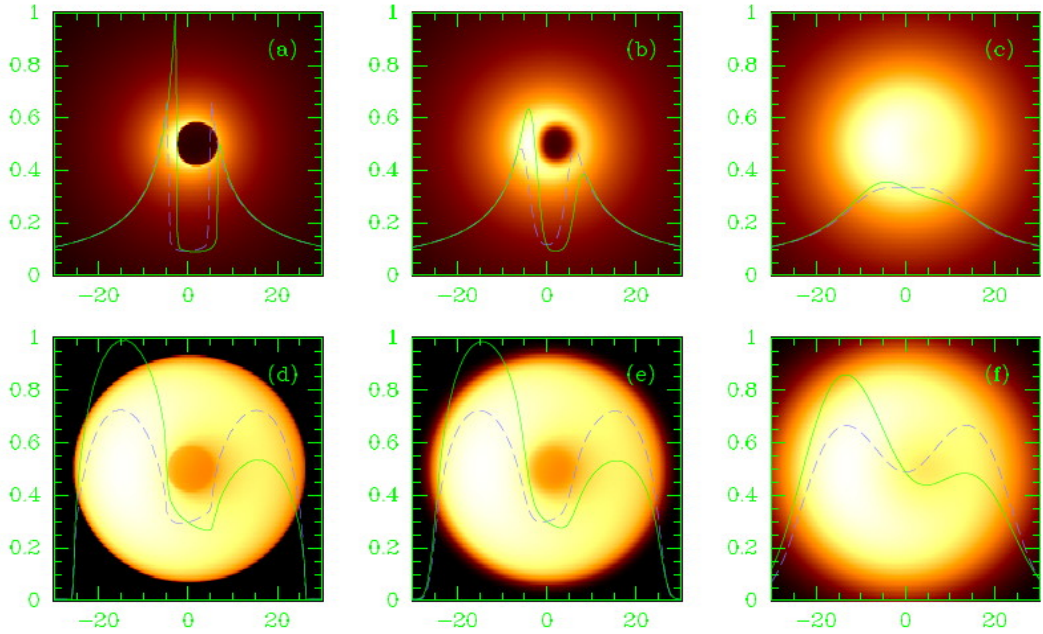

**Figure 22.** Images of an optically thin emission region surrounding the galactic black hole SgrA*. The black hole is maximally rotating (a = 0.998) in the top row and non-rotating in the bottom row. The emitting gas is assumed to be in free fall (top) or on Keplerian shells (bottom) with a viewing angle 45°. The left column shows the ray-tracing calculations in general relativity, the other columns are the images seen by an idealized VLBI array at 0.6 mm and 1.3 mm wavelengths, taking account of the interstellar scattering (from Falcke et al. 2000).

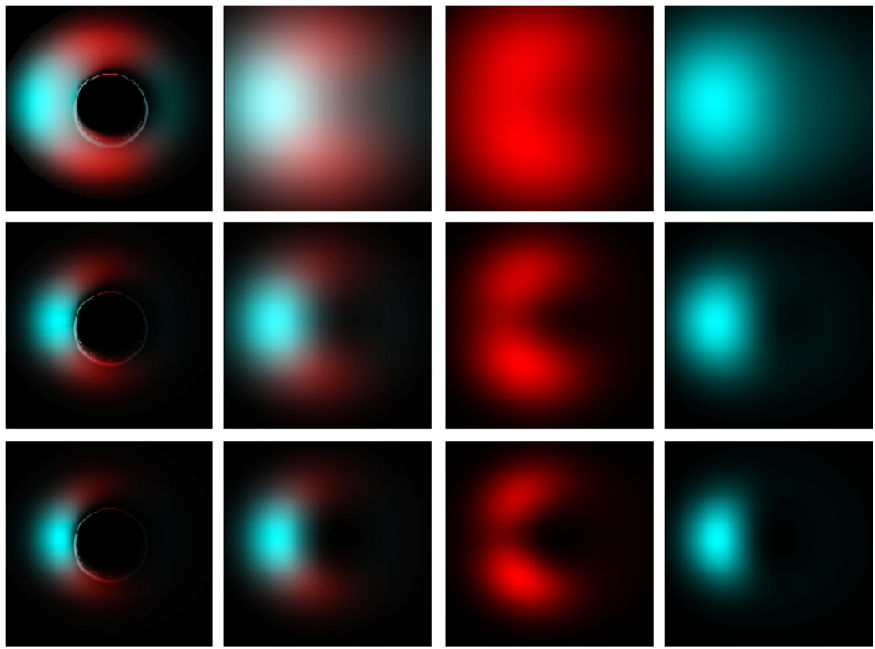

**Figure 23.** Polarization maps at three wavelengths (1.5 mm, 1 mm, 0.67 mm from top row to bottom row) calculated for the galactic black hole candidate SgrA*. The left most column shows how the radio maps might look seen from a close observer, the other columns show how the map might look from Earth with our vision blurred by gas in interstellar space (from Bromley et al. 2001).

On the observational side, successive radio imaging observations progressively reduced the size of emission region if SgrA*. A breakthrough was to extend VLBI to 1mm wavelength, where the scattering effects are greatly reduced and angular resolution is matched to the shadow of the galactic black hole. Then the collective effort was named the "Event Horizon Telescope" as the natural convergence of many historical and parallel works done by several independent teams in the world (Doeleman et al. 2009). The later measurement of the size of the 6 billion solar mass black hole in M87 gave a second source suitable for shadow imaging (Doeleman et al. 2012). Now the Event Horizon Telescope Consortium involves 20 universities, observatories, research institutions, government agencies and more than a hundred scientists who hope to make black hole imaging a reality as soon as 2019.

To conclude this survey, the path from idea to reality can take very a long time. Imaging black holes, first with computers, now with telescopes, is a fantastic adventure. Forty years ago I couldn't hope that a real image would be reachable in my lifetime and that, thanks to contributions by so many dedicated colleagues, my dream would become true.